
\magnification=\magstephalf
\hsize=16.0  true cm
\baselineskip=13pt
\parskip=0.2cm
\parindent=1cm
\raggedbottom

\def\pp{\parshape 2 0truecm 16truecm 1truecm 15truecm}
%
\def\singlespace{\baselineskip 12pt \lineskip 1pt \parskip 2pt plus 1 pt}

\def\doublespace{\baselineskip 24pt \lineskip 10pt \parskip 5pt plus 10 pt}

\def\apjref#1;#2;#3;#4 {\par\pp#1 \   #2,  #3, #4 \par}
%
%
\def\oldapjref#1;#2;#3;#4 {\par\pp#1, {\it #2}, {\bf #3}, #4. \par}
%
%

%
%
\def\subsection#1{\medskip\goodbreak\noindent\underbar{#1}}
\def\subsubsection#1{{\noindent \it #1}}
\def\ltsima{$\; \buildrel < \over \sim \;$}
\def\simlt{\lower.5ex\hbox{\ltsima}}
\def\gtsima{$\; \buildrel > \over \sim \;$}
\def\simgt{\lower.5ex\hbox{\gtsima}}
%
%



\def \eg {{\it e.g.\/}}

\def \etal {{ et al.\/}}

\def \cf {{\it cf.\/}}

\newcount\notenumber
\notenumber=1
\def\note#1{\footnote{$^{\the\notenumber}$}{#1}\global\advance\notenumber by 1}

\newcount\eqnumber
\eqnumber=1
\def\new{{\rm(\chaphead\the\eqnumber}\global\advance\eqnumber by 1}
\def\ref#1{\advance\eqnumber by -#1 (\chaphead\the\eqnumber
     \advance\eqnumber by #1 }
\def\last{\advance\eqnumber by -1 {\rm(\chaphead\the\eqnumber}\advance
     \eqnumber by 1}
\def\eq#1{\advance\eqnumber by -#1 equation (\chaphead\the\eqnumber
     \advance\eqnumber by #1}
\def\eqnam#1#2{\immediate\write1{\xdef\ #2{(\chaphead\the\eqnumber}}
    \xdef#1{(\chaphead\the\eqnumber}}

\newcount\fignumber
\fignumber=1
\def\nfig{\the\fignumber\ \global\advance\fignumber by 1}
\def\nfiga#1{\the\fignumber{#1}\global\advance\fignumber by 1}
\def\rfig#1{\advance\fignumber by -#1 \the\fignumber \advance\fignumber by #1}
\def\fignam#1#2{\immediate\write1{\xdef\
#2{\the\fignumber}}\xdef#1{\the\fignumber}}

\newbox\abstr
\def\abstract#1{\setbox\abstr=\vbox{\hsize 5.0truein{\par\noindent#1}}
    \centerline{ABSTRACT} \vskip12pt \hbox to \hsize{\hfill\box\abstr\hfill}}




\def\s{\ifmmode \widetilde \else \~\fi}
\def\={\overline}

\def\spose#1{\hbox to 0pt{#1\hss}}

\def\etal{{\it et al.\ }}
\def\cf{{\it cf.\ }}
\def\eg{{\it e.g.,\ }}

\def\lta{\mathrel{\spose{\lower 3pt\hbox{$\mathchar"218$}}
     \raise 2.0pt\hbox{$\mathchar"13C$}}}
\def\gta{\mathrel{\spose{\lower 3pt\hbox{$\mathchar"218$}}
     \raise 2.0pt\hbox{$\mathchar"13E$}}}
\def\Dt{\spose{\raise 1.5ex\hbox{\hskip3pt$\mathchar"201$}}}	
\def\dt{\spose{\raise 1.0ex\hbox{\hskip2pt$\mathchar"201$}}}	


\def\=={\equiv}

\def\dotsfill{\leaders\hbox to 1em{\hss.\hss}\hfill}









\rm		

\def\Atoday{\ifcase\month\or
  January\or February\or March\or April\or May\or June\or
  July\or August\or September\or October\or November\or December\fi
  \space\number\day, \number\year}
\def\Etoday{\number\day\space\ifcase\month\or
  January\or February\or March\or April\or May\or June\or
  July\or August\or September\or October\or November\or December\fi
  \space\number\year}

\def\apjref#1;#2;#3;#4;#5 {\par\pp#1 \   #3, #4, #5 (#2). \par}
%
%
  
  \newcount\fcount \fcount=0
  \def\ref#1{\global\advance\fcount by 1 \global\xdef#1{\relax\the\fcount}}
\tolerance 1500
\doublespace
\centerline { Dwarf Spheroidal Galaxies : Keystones of Galaxy Evolution}
\bigskip
\centerline {John S.~Gallagher, III}
\centerline {Department of Astronomy, 5534 Sterling Hall,}
\centerline {University of Wisconsin, 475 N. Charter St., Madison, WI 53706}
\centerline {Electronic Mail: jsg@jayg.astro.wisc.edu}
\bigskip
\centerline {Rosemary F.G.~Wyse}
\centerline {Department of Physics and Astronomy\footnote{$^1$}{Permanent
Address}}
\centerline {Johns Hopkins University}
\centerline {Baltimore, MD 21218}
\centerline {and}
\centerline {Center for Particle Astrophysics}
\centerline {301 Le Conte Hall}
\centerline {University of California}
\centerline {Berkeley, CA 94720}
\smallskip
\centerline {Electronic Mail: wyse@physics.berkeley.edu}
\bigskip
\bigskip
ABSTRACT: Dwarf spheroidal galaxies are the most insignificant extragalactic
stellar systems in terms of their visibility, but potentially very
significant in terms of their role in the formation and evolution of much
more luminous galaxies.  We discuss the present observational data and their
implications for theories of the formation and evolution of both dwarf and
giant galaxies. The putative dark matter content of these
low-surface-brightness  systems is of particular interest, as is their
chemical evolution.  Surveys for new dwarf spheroidals hidden behind the
stars of our Galaxy and those which are not bound to giant galaxies may give
new clues as to the origins of this unique class of galaxy.

\bigskip
\centerline{\bf 1. INTRODUCTION}

The astrophysical significance of small extragalactic systems, the dwarf and
related classes of galaxies, remains unclear (Hodge 1971). Did these
galaxies once lie along the main path of galaxy formation, or are they
merely collections of leftover matter produced by secondary processes?  Are
they strongly bound, or are they dissolving in the Galactic tidal field?
(Hodge 1964; Hodge and Michie 1969).  Cold dark matter (CDM) models for
galaxy formation have served to focus these issues.  In CDM low mass
galaxies would be the first objects to become bound and drop out of the
expanding Universe (Blumenthal et al. 1984).

In general, these small systems should not have survived to the present day,
but should be subsumed into larger systems (White and Rees, 1978; Moore,
Frenk and White 1993), or, more correctly, their dissipationless dark halos
should have merged, with the fate of the visible galaxy being very uncertain
(Lacey and Cole, 1993; Kaufmann and White, 1993). The baryonic material will
keep its separate identity only if it has dissipated enough energy to be
more tightly bound than the next level of the hierarchy (White 1983; Cen and
Ostriker 1993); this may indeed be the case (Navarro \etal\ 1994). The ideas
of the CDM model lead naturally to the question of whether
dark-matter-dominated, low-mass galaxies exist in nature, and to the
build-up of luminous galaxies like the Milky Way by merging of substructure
(e.g. reviewed by Silk and Wyse 1993).  Extracting the merging history --
and future -- of the Milky Way Galaxy may be possible by study of satellite
galaxies.

As a result there have been renewed efforts to measure the spatial
distributions, dynamics, structures, and other properties of low-mass
galaxies in a variety of environments (Faber and Lin 1983; Lin and Faber
1983; Thuan, Gott and Schneider, 1987; Eder et al. 1989; Puche and Carignan,
1991; Vader and Sandage 1991; Ferguson and Sandage 1990), including
low-surface-brightness galaxies (Disney and Phillips 1987; Impey, Bothun and
Malin 1988; Schombert et al. 1992; Sandage, Binggeli and Tamman, 1985;
Karachentseva, Schmidt and Richter 1984). The advent of automated
plate-measuring machines and large format CCDs coupled to multi-object
spectrographs has resulted in data for large numbers of individual stars in
several of the Galactic dwarf spheroidal (dSph) systems, together with
rigorous searches for more companions to our Galaxy.

 Satellite galaxies  at high Galactic latitude may often be found by simply
plotting stellar surface density contours projected on the sky. For example,
the dwarf galaxy discovered in the Sextans constellation, latitude $b =
42.3$, by Irwin et al. (1990) is clearly seen in their Figure 1(b), which
plots all stellar objects down to the plate limit ($B_J \sim 22.5$); note
the paucity of objects classified as stars -- the contours plotted start at
1.5 images/arcmin$^2$ and increase in steps of 0.5 images/arcmin$^2$. The
dwarf has an angular extent of a degree or so. Dwarf galaxies could be
hiding at lower latitudes, where there is sufficient foreground disk that
the over-density of the satellite is too small a perturbation to be detected
by a simple sum of all stellar images.  An example of this is the
Sagittarius dwarf, recently discovered (Ibata, Gilmore and Irwin 1994), as
part of a survey of the kinematics of the bulge (Gilmore and Ibata, in
prep).  The dwarf initially manifest itself as a moving group, with very
distinct kinematics from the bulge, and rather localised in angular
position.  The color-magnitude diagram of the field where the moving group
is seen kinematically contains a fairly obvious core-helium-burning red
`clump', indicative of an intermediate-age population, and giant branch,
which are not detected in those  fields where the stars have normal bulge
kinematics. Isolating stellar images which have the  color and apparent
magnitude of the clump, and plotting isodensity contours of the difference
between the `moving group' field and an offset field, revealed the
spatially-extended dwarf galaxy (Ibata, Gilmore and Irwin 1994). The
isophotes are very aspherical, suggestive of tidal stresses. The distance to
the dwarf may be estimated from the clump/HB magnitude, resulting in a
position $\sim 12$ kpc from the Galactic center on the other side of the
Galaxy, $\sim 5$ kpc below the disk plane.  The dwarf is inferred to be
rather massive, similar to the Fornax dwarf galaxy, from application of the
observed luminosity-metallicity relation of the extant satellite galaxies
(Caldwell et al. 1992), and from the red clump/giant population. At least
three globular clusters have kinematics and distances that are suggestive of
their being physically associated with the dwarf galaxy. Follow-up deep
photometry has confirmed the dSph nature of the Sagittarius dwarf and the
intermediate age, $\sim 10$~Gyr (Mateo et al.~1994).

Multi-object spectroscopy on large telescopes has provided radial velocities
and chemical abundances for large enough samples of stars in the Galactic
dSph to begin to make quantitative analyses of the internal dynamics and
enrichment histories, while deep, large-area color-magnitude diagrams allow
the dSph stellar populations to be disentangled from the foreground Galaxy.
Proper motions of the Galactic-companion dSph are beginning to become
available, and will constrain the orbits of these objects. Searches for
low-surface-brightness dwarfs beyond the Local Group using large-format CCDs
should reveal counterparts to the local dSph; large-area photometric surveys
such as the Sloan Digital Sky Survey will provide ideal datasets for
identification of dSph. This review aims to put these new and anticipated
results in context.

The least-massive known galaxies are the dSph systems.  The dSph galaxies
exist at a key position in the hierarchy of spheroidal stellar systems --
although as we discuss further below, whether or not they are bound is open
to question.  In some properties they appear to be the logical extension of
the diffuse dE galaxy structural family to low luminosities (Wirth and
Gallagher 1984; Kormendy 1985), and yet in some respects also resemble
low-luminosity disk galaxies (Kormendy 1987). Additionally, they overlap in
luminosity with the more massive globular star clusters such as 47 Tuc or
Omega Centauri (Webbink 1985).

However, despite their similarities in total stellar luminosities, dSph
galaxies and globular clusters have dramatically different structures (de
Carvalho and Djorgovski 1992). Globulars are centrally concentrated and have
central stellar densities, even in the relatively unconcentrated Palomar
globulars, that are $>$ 100 times the central stellar densities of the much
more diffuse dSph galaxies (Webbink 1985; Djorgovski 1993).  A wide-ield
image of the Ursa Minor dSph is shown in Figure 1.  Indeed, our current
understanding of star formation suggests that the stars in dSph could not
have formed at the currently observed low stellar densities. Evidently
something happened between the epoch of star formation and now, a
`something' that is most probably related to the inferred low escape
velocities of these diffuse systems (Dekel and Silk 1986). The presence of
globular star clusters in the Fornax dSph demonstrates that at least this
dSph is a true galaxy, and extends the trend for all known globular star
clusters to be bound to galaxies (although the search for truly
intergalactic globular clusters  continues; Arp and Madore 1979).  The
significance of the fact that the Fornax dwarf has many more globular
clusters per unit galaxy luminosity than expected from normal galaxies is
not clear, although perhaps the lower tidal field and lack of other internal
destruction mechanisms could be a factor (Rodgers and Roberts 1994).

Equilibrium dynamical models which require large dark-matter densities have
now been constructed for several dSph galaxies, being most extreme (in terms
of mass-to-light ratio), for the Draco and Ursa Minor galaxies (Lake 1990a;
Pryor and Kormendy, 1990). In an insightful pair of papers Faber and Lin
(1983) and Lin and Faber (1983) suggested that the presence of a dominant
dark-matter component is the key feature which distinguishes galaxies from
star clusters (luminous mass cannot be the crucial factor) and that all
small galaxies have similar and large dark-matter components.  Since the
publication of these papers evidence has been mounting for dark matter in
the dwarf {\it irregular\/} galaxies,  with the trend that lower luminosity
disk systems (Sc and later) have relatively more dark matter (Puche and
Carignan 1991; Ashman, Salucci and Persic 1993). It is tempting to identify
the dSph galaxies as the gas-poor counterparts of irregulars, but many
questions remain, some of which we address in this review.

A summary of the currently-known galaxies in the Local Group is given in
Table 1 (based on van den Bergh 1992). The membership of the Local Group
also becomes uncertain at larger distances, e.g., for galaxies like Leo A
(see van den Bergh 1994a; Hoessel et al. 1994).

We define a dSph to be a galaxy with a total luminosity of $M_B > -14$, low
optical surface-brightness (fainter than 22 V-magnitudes per square arcsec)
and no nucleus.  Note that this definition excludes NGC 205, NGC 147, NGC
185 and many of the elliptical dwarfs found in the Virgo cluster, even those
of low surface-brightness (Binggeli and Cameron 1991). In particular, {\it
bona fide\/} dSph galaxies, and certainly all dSph systems for which we will
have good dynamical models in the near future, are located near the Milky
Way and M31, two massive spirals.  The close association between the dSph
dwarfs and massive galaxies has a number of significant consequences which
have been individually discussed in the literature during the past decade.
In this paper we will (1) revisit the implications of the satellite status
of Galactic dSph systems, especially in terms of the assumption of dynamical
equilibrium; (2) consider why such connections might occur between dwarf and
giant galaxies; and (3) emphasize the importance of resolving these
questions.

\centerline{\bf 2. BASIC PROPERTIES OF dSph GALAXIES}

Recent papers by, among others, Da Costa (1992),  Caldwell et al. (1992),
Mateo \etal\ (1993), Zinn (1993), and Kormendy and Bender (1994) provide
good overviews of dSph galaxies.  Properties of the Galactic dSph systems
can be divided into three main areas consisting of structural parameters
(mass, size, internal kinematics), stellar populations (ages, chemical
elemental abundances, binary star fractions, gas content), and environment
factors (orbits, correlations between internal properties and
Galacto-centric radius).  A summary of basic dSph characteristics drawn from
the current literature (Hodge 1982; Zaritsky et al. 1989; Demers and Irwin
1993; Lehnert et al. 1992) is given in Table 2.

{2.1 Structure}

The dSph are apparently ellipsoidal galaxies of low central surface
brightness.   Their stellar density profiles, like those of other dE
systems, are fit either with low central-concentration King models or with
exponential disks having scale lengths of 100 - 400 pc (see Figure 2).  The
outer radii of these systems are difficult to determine due to their large
angular sizes and faintnesses.  Visual central surface-brightnesses are in
the range of 22 -- 26 mag/arcsec$^2$ and are not strongly correlated with
the absolute visual magnitudes, which extend down to M$_V = -9$ for the Ursa
Minor system (Lake 1990a; Mateo et al. 1993; Nieto et al. 1990).

The dSph, like other dEs, are in one way more \lq\lq disky" than even
spirals.  Where van der Kruit and Searle (1982) have found the exponential
stellar disks of spirals often end rather abruptly after about 5 scale
lengths, the Andromeda dSph remain exponential in form out to nearly seven
scale-lengths, where they are lost into the sky background (Armandroff et
al. 1993). The presence of such smooth edges in dSph and dE galaxies are not
understood, but could be a symptom of the effects of tidal heating (Aguilar
and White 1986).  However, it should be remembered that perhaps as many as
ten orbits are required before a tidal radius is well-defined, with no
sustained memory of the initial radius (Seitzer 1985), and the dSph are
sufficiently distant that this is a substantial fraction of a Hubble time.
The (projected) axial ratios of the dSph are in the range of $0 \simlt (1 -
b/a) \simlt 0.6 $ (Hodge 1971, Caldwell et al. 1992), and are not very
different from the range seen in dwarf irregular galaxies (Lin \& Faber
1983; note that the axial ratios of the dSph may be due in part to tidal
stretching, as discussed below).

Internal kinematics are now available for most of the known Galactic dSph
galaxies from high precision observations of radial velocities of individual
stars, with more data being analysed. Several points stand out from the
current kinematic data (Kormendy 1985; Pryor and Kormendy 1990; Ashman 1992;
Freeman 1993; Suntzeff et al. 1993; Hargreaves et al. 1994a,b):

(1) The central velocity data can be fit by a simple Gaussian
projected-velocity distribution function, but it is not yet established that
this distribution is independent of location within the galaxy, though this
is the model generally used.   The spatial variation of the internal stellar
kinematics is of great importance in constraining dynamical models, but the
data analysed to date have not allowed any strong conclusions; data which
several research groups have in hand should be capable of distinguishing
dark-matter models.

(2) All of the dSph central velocity dispersions are in the 5-12 km/s range.
 This is near the magic velocity dispersion of young stars and HI gas in the
thin disks of spirals (van der Kruit and Shostak 1984), generally believed
to have something to do with hydrogen cooling at T$< 10,000$ K, and to the
stellar velocity dispersions in globular star clusters (Illingworth 1976),
but is larger than the velocity dispersions of typical OB associations. The
velocity dispersions also do not correlate with optical luminosities.  If
interpreted using the virial theorem, the conclusion is that the dSph all
have approximately the same mass, while their luminosities span a wide
range.

(3) There is no evidence for rotational support to very low levels in some
of the dSph for which suitable spaially-resolved kinematic data are
available, such that $v/\sigma < 0.3 $ in the Fornax (Freeman 1993) and
Sculptor dSph systems. In this regard the dSph, although appearing
moderately flattened, are like other dE galaxies in being supported by
anisotropic velocities (Bender and Nieto 1990; Mateo et al. 1991). A
shear-velocity field has been detected in the Ursa Minor dSph (Hargreaves et
al. 1994b), consistent with rotation about the apparent major axis, with
$v/\sigma$ of order unity in the outer regions.  These data are interpreted
as the result of tidal stretching along the initial minor axis combined with
the generation of a net azimuthal streaming by preferential tidal stripping
of stars on prograde orbits (Hargreaves et al. 1994b). Spatially-resolved
kinematics will allow better equilibrium dynamical models to be constructed
for dSph galaxies, as has been done for globular clusters (Lupton et al.
1989; Gunn and Griffin 1979).

The density structures of dSph galaxies have been widely discussed in the
literature, especially in connection with the `dark-matter problem'.  The
simplest approach for finding core M/L ratios is to fit core velocity
dispersion data, together with the star-count brightness profiles, to
equilibrium dynamical models with isotropic velocity dispersions (Lake
1990a; Pryor and Kormendy 1990; Hargreaves et al. 1994a). This process
yields M/L values in the range from 5-15 for the Fornax and Sculptor dSph
systems and up to 100 for the Ursa Minor, Sextans, and Draco systems.  These
models applied to Ursa Minor, Carina, and Draco also give high central
densities, of about $0.1 M_\odot/pc^{3}$, most of which must be in the form
of dark matter (Lin and Faber 1983). This contrasts with inferred central
dark-matter densities of $\simlt 0.01 M_\odot/pc^{3}$ for gas-rich dwarf
irregulars (Puche and Carignan 1990) and typical spiral galaxies (Kent
1987). The lack of rotational support to some of these dSph however implies
{\it anisotropic\/} velocity dispersions; as shown by Pryor and Kormendy
(1990), use of anisotropic models would result in a somewhat reduced core
mass density.  The uncertainties and subtleties associated with the
determination of M/L are discussed, for example, by Hargreaves et al.
(1994a,b), and in Binney and Tremaine (1987).

Lack of knowledge about orbits also complicates discussions of the internal
dynamics of the dSph. If we assume that the Galactic dSph are on only
moderately elliptical orbits, then the orbital periods are several times
longer than the internal crossing times. Alternatively, if these dSph are
now making their apo-Galacticon passages on highly eccentric orbits, then
since the time spent in the dense inner Galaxy is a small fraction of the
orbital period, the time scale for variations in the tidal field produced by
the Milky Way may be close to the internal crossing time. Thus the question
of the degree to which dSph galaxies are in dynamical equilibrium remains
open. The ongoing efforts to measure proper motions of the nearest dSph
galaxies (Scholz and Irwin 1993; Majewski and Cudworth 1993; Cudworth,
Schweitzer and Majewski 1994), which will allow three-dimensional orbits to
be determined, will be a good test of models.

{2.2 Stellar Populations}

The dSph galaxies occupy an intermediate position in terms of stellar
populations.  Their metal abundances as inferred from the spectra and colors
of red giants fall in the regime defined by Galactic globular clusters, with
most Galactic dSph galaxies having mean metallicities similar to the peak of
the halo globular clusters, or $-1.5$ dex. An example of a color-magnitude
diagram for the Carina dSph is shown in Figure 3, taken from Smecker-Hane
\etal\ (1994).

However dSph galaxies are unlike most globulars in that a range of
abundances is present in several of the Galactic dSphs (see Table 2).
Further, the globular clusters of the Fornax dwarf do not all have the same
chemical abundance, but  range from $-1.35 $ dex to $-1.93 $ dex (Dubath,
Meylan and Mayer 1992). In addition the Fornax, Carina, Draco, and Leo I
dSphs have intermediate-age stellar populations, identified from AGB stars
and a tentative turnoff population, with an inferred last epoch of major
star formation of near 5 Gyr (Aaronson 1986; Mighell 1990; Mighell and
Butcher 1992; Lee et al. 1993; Smecker-Hane et al. 1994; reviewed by
Freedman 1994). These four dSph systems therefore are fundamentally
different from globular star clusters, and it is likely that age spreads
also exist in other dSphs, but these are not trivially detected in faint old
stellar populations where the main sequence turnoff is difficult to identify
(Freedman 1994). Indeed, the stellar population of Leo I is predominantly of
intermediate age, with no obvious signature of very old stars (Demers, Irwin
and Gambu 1994).  There may be a trend of the percentage of intermediate-age
stars with distance from the Milky Way, such that the more distant the dSph,
the larger the fraction of younger stars (Silk, Wyse and Shields 1987; van
den Bergh 1994b), although the newest Galactic dSph, the Sagittarius dwarf,
does not fit this trend.

The evolution of gas in small galaxies is probably largely controlled by the
flow of matter into and out of the systems.  Due to the low escape
velocities of dSph galaxies, the ability of these systems to retain SNe
ejecta is limited (Larson 1974; Vader 1986; Dekel and Silk 1986) and thus
metallicities will depend on mass loss perhaps even more than on the star
formation history (Sandage 1965; Vader 1986; De Young and Gallagher 1990).
Furthermore, the internal dynamical evolution of the system may be greatly
affected by galactic winds. Meurer et al. (1992) have found compelling
evidence for star-formation-driven gas loss from a dwarf galaxy; ongoing
surveys (Marlowe 1995) will allow an estimate of the importance of galactic
winds in general. In addition small galaxies may be able to capture cooling
gas from the circumgalactic environment, allowing rejuvenation of the
stellar population (Silk, Wyse and Shields 1987) -- thus competition for gas
could provide an explanation for trend of intermediate age population with
distance from the Milky Way. This process will further modify metallicity
distributions, with possible signatures in chemical element ratios -- those
elements primarily synthesised in Type II supernovae, such as oxygen, may be
affected more than those elements primarily synthesised in the deaths of
long-lived progenitor stars, such as iron (Wheeler, Sneden and Truran 1989;
Gilmore and Wyse 1991).

The galactic-wind model for Galactic dSph also seems to be consistent with
the lack of any detectable interstellar matter in these systems. Searches
for cool gas via the HI 21-cm line have yielded only upper limits (Knapp,
Kerr and Bowers, 1978; Mould, Stavely-Smith and Wright 1990), with
$M_{HI}/L_V \simlt$ 0.01, although the confusing effects of the Magellanic
Stream and high velocity HI complicate searches for cool gas in nearby
galaxies. This should be compared with the $M_{HI}/L_V$ ratios of 0.01-0.1
observed for some dE and transition dSph/dIrr galaxies, such as the Phoenix
dwarf (Carignan, Demers and C\^ot\'e, 1991). As a group the Galactic dSph
galaxies are among the most HI-poor galaxies known.

An examination of Table 2 suggests the complexities expected from
galactic-wind models are present  in chemical abundances of the Galactic
dSph retinue.  The most massive systems have higher metallicities as
predicted by wind-induced mass-loss models, but the presence of an
intermediate-age stellar population in Carina indicates that other effects
have also been important.  In a wind mass-loss model we would expect that
the presence of intermediate-age stellar populations, the metallicity levels
and the galactic mass correlate, and yet this does not appear to be strictly
true.  However, the low mean  metallicity of the stars in dSph galaxies is
indeed the expected signature for galaxies where chemical evolution has been
strongly influence by gas loss via winds rather than by astration, assuming
an invariant stellar Initial Mass Function (IMF; Hartwick 1976). In
addition, the distribution of stellar metallicities in a few dSph appear to
be skewed as expected if sudden loss of gas truncated the chemical evolution
(Suntzeff 1993).

{2.3 Orbits and Environment}

The orbits of the dSph galaxies will not be fully known until proper motion
measurements begin to become available within the next few years.  These
measurements are extremely difficult because the proper motions are expected
to be small. For example, a transverse velocity of 100 km s$^{-1}$ at a
distance of 50 kpc yields a proper motion of about $\mu =$0.5 milliarcsec
yr$^{-1}$. This is currently just feasible by using extragalactic objects to
define the stationary reference background on deep plates taken with large
reflectors over multi-decade time baselines (see Majweski \& Cudworth 1993;
Kroupa, R\"oser, \& Bastian 1994).

Even in the absence of detailed orbits, the available radial velocities,
however, already present an interesting picture (Zaritsky et al. 1989). If
the dSph were on circular orbits then their Galactocentric radial velocities
should be a small fraction of the circular velocity at their locations.  For
example, if we adopt a maximal mass model for the Galaxy with $V_c=220$ km/s
at 100 kpc, we would expect $V_r \simlt 25$ km/s for near circular orbits:
thus we immediately see that most of the Galactic dSph cannot have circular
orbits. Little and Tremaine (1987) and Fich and Tremaine (1991) analyzed the
dynamics of dSph systems in the context of determining the Galactic mass
distribution.  They conclude that most of the nearby dSph are likely to be
bound to the Milky Way and have phase-mixed over their orbits, which have a
range in eccentricities.  Thus it is proper to assume that most dSph
galaxies have completed several orbits around the Milky Way since their
formation, and with the possible exception of Leo I (Byrd et al. 1994), are
not newly acquired members of the Galactic family.

Three additional interesting points emerge from these discussions: (1) With
the exception of the disturbed Saggitarius dSph, the innermost Galactic dSph
are currently located near the estimated outer radius of the Galactic dark
halo, $R \simgt 50$ kpc. (2) The orbits of some dSph may be sufficiently
extended that the influence of M31 needs to be taken into account. (3) In
the absence of proper motion measurements, we cannot prove that the dSph are
bound to the Milky Way.
\medskip

\centerline{\bf 3. THE PRICE OF PROXIMITY: EFFECTS OF THE MILKY WAY}

The array of Galactic dSph galaxies presents us with a highly ordered
distribution across the sky.  As noted several years ago (Kunkel and Demers
1976; Lynden-Bell 1982a,b) most of the dSph systems fall near either the
great circle defined by the Magellanic Stream or a second, nearby great
circle which perhaps associated with the Fornax system.  The Galactic dSph
retinue does not have the appearance expected if these systems were being
captured from a postulated reservoir of faint dwarf systems (Tremaine 1987,
Majewski 1994). If we want to make dSph galaxies as independent entities,
then a plausible physical mechanism must be found to order these systems
into the present, near polar-ring structure about the Milky Way.

One might appeal to dynamical friction to accomplish the task of selecting
which Galactic satellites could survive. However, even for a dark-matter
dominated system with $(M/L)_V=100$, the dynamical friction time scales can
be estimated from standard relationships (\eg\ Binney and Tremaine 1987)  to
give
$$t_{df}  \sim (M_{gal}/M_{satellite}) t_{crossing} > 10^{12}{\rm
yr},$$
for circular orbits at the present radii of the inner Galactic dSph.  The
masses of existing dSph galaxies are too low by a factor of 100 for
dynamical friction to be important at their present locations.  And even if
the orbits were eccentric, an equivalent circular radius of $<10$~kpc is
needed for these systems to experience orbital decay in less than 10$^{10}$
yr.  The orbits of the well-studied dSph galaxies around the Milky Way have
yet to be profoundly influenced by dynamical friction (Quinn and Goodman
1986). As a corollary, there appears to be no obvious way for dynamical
forces associated with a spheroidal Galactic dark halo to have aligned an
initially random distribution of dSph galaxies into the current
near-polar-orbit array.

A second destruction mechanism for satellites of galaxies is tidal heating
and disruption.  In its simplest form this process can be viewed as the
shrinking of a zero-velocity surface about a mass distribution.  A rough
guide to the tidal radius $r_j$ is then given by the Jacobi limit (Binney
and Tremaine 1987), essentially that disruption occurs when the mean density
of the satellite and parent galaxy are equal. This can be written in a
useful form for the dSph systems as

$$ r_j= 1.5[(L_V(dSph)/10^7
L_\odot)(10^{12}M_\odot/M_{MW})(M/L_V)_{dSph}]^{1/3}
     (D/100 {\rm kpc) kpc. }$$

Thus for the Draco dwarf with $L_V=2\times 10^5 L_\odot$, D=75 kpc, the
tidal radius is $r_j = 300 (M/L_V)^{1/3}$ pc, or $r_j < 600 $ pc for a
normal stellar composition.  While the radii of dSph galaxies are extremely
difficult to determine observationally, Lake (1990a) suggests that the
photometric radii based on the Hodge star counts for the Draco and Ursa
Minor dSphs are $r_* > 25 $ arcmin, or $>0.5$ kpc. Thus the sizes of the
dSph are at least close to the expectations of a simple tidal model,
although their internal velocity dispersions suggest that they are actually
too tightly bound for tidal shredding to be effective.

The real situation is certainly even more complex than indicated by
this rough calculation.  For example, Kuhn and Miller (1989)
suggest that resonant coupling may enhance tidal heating and also
lead to the disruption of dSph galaxies.  More extensive
discussions of the complex dynamical issues associated with tidal
interactions can also be found in the literature describing the
destruction of globular star clusters (e.g., Aguilar 1993, and
references therein). We also note that the qualitatively expected
effects of tidal heating are clearly present observationally in
outer regions of some elliptical galaxies (Kormendy 1977).

Another tidal limit may be set if dSph get sufficiently close to the Milky
Way to interact with the disk.  This could produce disk shocking which will
influence the internal structures of dSph when their densities are
comparable to that of the local disk (Aguilar 1993; Capaciolli, Piotto and
Stiavelli, 1993).  Additionally, passages through the outer, gas-rich
Galactic disk would lead to an external ram pressure on gas within dSph,
especially if the disk extends in ionized gas beyond the neutral HI edge
(Maloney 1992). This mechanism would then further reduce the ability of
these systems to retain gas. Such an effect may be operative on the
Magellanic Clouds, producing the Magellanic Stream (Mathewson et al. 1987;
Wayte 1991), although a straightforward tidal origin for the Stream remains
viable (Gardiner, Sawa and Fujimoto 1994). Note that both bound and unbound
orbits of the LMC are consistent with the present proper motion measurements
(Kroupa, R\"oser and Bastian 1994).

With the recognition of tidal heating as a destructive mechanism for dSph
galaxies, we can divide models of the internal dynamics of dSph systems into
two major classes. The equilibrium models deal with each galaxy individually
and assume that external processes have had a negligible effect on the
internal structures.  This class of model yields high dark-matter densities
for some dSph galaxies, including the Carina system which is currently too
far from the Milky Way to experience significant tidal effects (Pryor and
Kormendy 1990). In this picture a universal structure for the dSph would be
that of dense galaxies which are gravitationally equivalent to rocks and are
not influenced by the Galactic tides.  The spatial distribution of dSph
galaxies then is in roughly a steady state and will evolve only slowly in
response to dynamical perturbations (e.g., passages of the Magellanic
Clouds).

An alternative view is that the dSph are as fragile as they appear from
their low stellar surface densities. These systems then are being tidally
heated and thus eventually will be destroyed (Kuhn and Miller 1989; Kuhn
1993). In this model the Galactic population of dSph galaxies is evolving,
and some systems could have been disrupted in the past.  Indeed we would
expect that the innermost mean radius where dSph systems are found might
move outward with time as closer-in objects are disrupted.  The present-day
dSph galaxies then would be remnants of what may once have been a more
extensive satellite population.

\bigskip
\centerline{\bf 4. CONSEQUENCES OF `THE DARK-MATTER PROBLEM'}

The central issue in interpretation of the velocity dispersion observations
is whether or not the dSph galaxies are in dynamical equilibrium states. If
they are equilibrium systems, then a dominant dark-matter component is
required in several dSph systems. Furthermore, this dark component is
inferred to have central densities that are ten times larger than those
derived for more luminous galaxies. The nature of the dark matter is also an
open issue at present, with baryonic dark matter perhaps favored by the high
densities, which are suggestive of dissipative material. However, should the
dSph be in the process of disruption, then obviously the virial theorem
should not be applied, and the presence of an excess density of dark matter
is not secure.

We now consider the implications of the dark-matter problem by
considering three possible cases:

4.1  Non-Baryonic Dark Matter

Faber and Lin (1983) used equilibrium dynamical models to conclude that the
dSph probably have very high mass-to-light ratios, and therefore contain
large amounts of dark matter. This was later put in the cosmological context
of the hierarchical clustering Cold Dark Matter scenario, where the first
scales to turn-around, and presumably form stars,  are of the mass scale of
dwarf galaxies (Blumenthal et al. 1989). The properties of these systems and
their possible identification with present-day dwarf galaxies is discussed
in Dekel and Silk (1986). The persistence of dwarfs in groups of galaxies is
unclear -- the CDM halos of typical, i.e. `$1 \sigma$', perturbations on
this scale most certainly merge, but the fate of the luminous components is
unknown (Lacey and Cole 1993; Kauffman and White 1993).

However, the large inferred central dark-matter densities of the dSph
suggest that in this picture, the dSph evolved from higher amplitude density
fluctuations, greater than $3\sigma$, which collapsed and virialised at high
redshift, $1 +z \sim 20$ (\cf\ Evrard 1989). The dSph would then be the most
ancient galaxies. They are also relatively close to the Galaxy,  and highly
clustered, which may be consistent the statistics of density peaks (Kaiser
1984; Bardeen et al. 1986). The very dense individual dark halos of the dSph
would retain their identity even {\it within} the outer dark matter halo of
the Milky Way.  Thus this model ($3\sigma$ dwarfs in CDM) is dynamically
self-consistent. The low stellar densities would result from galactic winds.
In this model stars are born at normal densities and then expand in response
to mass loss driven by winds and supernovae from massive stars, although the
details remain to be calculated. In particular it is very hard to understand
why the dissipationless stellar population should have expanded in response
to gas loss while the dark matter remained dense.

4.2  Baryonic Dark Matter

Alternatively, the high densities inferred for the dSph dark matter could be
an indication of dissipation, consistent with the dark matter being baryonic
(Lake 1990b). The recent identification of `gold-plated' candidate
micro-lensing events in lines-of-sight towards to the Magellanic Clouds,
interpreted as lensing of background stars in the Clouds by foreground dark
objects in our Galaxy (Alcock et al. 1993; Aubourg et al. 1993), consistent
with lensing by objects with sub-solar mass, has revitalised interest in
brown dwarfs as halo dark matter.  However, this would require that the
stellar IMF at the time of dark halo formation were very different from the
present-day IMF, and indeed different from the IMF of the stellar halo which
appears to show a downturn below a few tenths of a solar mass, much like the
disk (Dahn 1994).  	It may be that the lensing events observed are more
like those seen towards the Galactic bulge (Udalski et al. 1993), which are
believed due to normal low mass stars, and that both lens and source are in
the Clouds. Further, deep K-band imaging of high latitude fields has failed
to find evidence for faint red stellar objects (Hu et al.~1994), as have
observations with Hubble Space Telescope (Bahcall et al.~1994). Baryonic
dark matter playing a dominant role in dSph would have the further
requirement that the IMF be variable from dSph to dSph.

Larson (1987) has suggested that an IMF biased towards massive stars could
self-consistently provide dark remnants and also provide self-enrichment to
give a correlation between M/L and [Fe/H], which has an uncertain
observational status in the Galactic dSph. However, as above, quite how the
dark matter remains dense while the stars become diffuse is unclear.  This
problem is exacerbated by the need to prevent too much enrichment from the
massive stars, which requires very significant gas loss, many times that of
the mass left behind.  This must occur on a timescale which is rapid
compared to the star formation rate, but slow compared to the internal
crossing time, to avoid unbinding the system.  Thus dSph with no evidence
for extended star formation are particularly difficult in this scenario.

A skew in the IMF (rather than truncation, which would further excacerbate
potential problems with unbinding the system) should be detectable from the
value of the ratios of alpha-capture elements  to  iron peak elements from
Type II supernovae (Wyse and Gilmore 1992). This effect reflects the
variation of oxygen and other nucleosynthetic yields with supernova
progenitor masses. For example, a slope with power law index flatter by
unity from the IMF in the solar neighborhood produces enhancements in oxygen
relative to iron that is perhaps a factor of three higher than that seen
(already a factor of three above solar) in the halo stars in the solar
neighborhood (Wyse and Gilmore 1992).

Note that in the lowest mass gas-rich galaxies for which good rotation curve
data exist, there is an apparent relationship between the gravitational
field and the HI gas.  This manifests itself  in the sense that the observed
rotation curve is a scaled version of the rotation curve one would derive
from the HI alone (Puche and Carignan 1991). A similar effect is also seen
in the outer parts of massive galaxies (Bosma 1982). One is tempted to
conclude that, like the HI, the halo dark matter is indeed baryonic
(Gallagher 1990), consistent with indications from Big Bang Nucleosynthesis
that a large fraction of baryons are dark (Walker et al. 1991), and indeed
just the amount of dark matter inferred on the scale of galactic halos.
However, it should be noted that the recent detection of unexpectedly large
amounts of deuterium in a quasar absorption line system by Songaila et al.
(1994) and by Carswell et al. (1994) would, if confirmed by observations of
other systems, make ubiquitous baryonic dark-matter haloes impossible.

4.3 No Dark Matter (NDM)

The possibility that dSph do {\it not\/} contain the dark matter inferred
from the analyses of the velocity data could arise in two ways. Firstly, the
internal orbital motions of the expected population of binary stars -- found
to be ubiquitous, at the 30\% level, in all normal stellar systems -- will
cause a zero point offset in the observed velocity dispersion. The
long-period binaries could provide a contribution of a few km/s to the
dispersion, which is very difficult to remove compared to the more obvious
higher orbital motion outliers. However, this is uncomfortably close to the
amplitude required to remove the need for dark matter, once subtracted off
from the observed velocity distribution to obtain the center of mass motions
of the binaries in the potential well of the dSph. Indeed, Suntzeff et al.
{}~(1993) find that a binary fraction of 25\% in giant stars, with periods
between 1yr and 1,000 yr, is sufficient to increase the velocity dispersion
to observed values even given a true M/L of only 3, as seen in normal old
stellar populations. This effect needs careful modelling (Mathieu 1985;
Suntzeff et al. 1993; Hargreaves et al. 1994).

Secondly, it is not firmly established that the dSph with very high inferred
M/L ratios are in virial equilibrium.  In this case there may be no dark
matter (NDM), and the dissolution timescale is the critical parameter -- it
cannot be too short!  For example, in this interpretation present
observations would suggest that 3/9 of the dSph companions to the Milky Way
would be in the process of coming apart, presumably due to tidal stresses,
and so without a steady source of dSph, we require that the dissolution time
be greater than 25\% of the age of the stellar populations of the dSph -- or
$\simgt 5$~Gyr. Furthermore, the association of apparently high M/L values
with the {\it cores\/} of dSph galaxies exacerbates this difficulty. While
these problems are somewhat ameliorated if some dSph are not bound to the
Galaxy, the problem of supply of dSph remains. Also unanswered in this model
is the origins of marginally-bound galaxies with ultra-low stellar
densities.

The NDM model is in principle feasible because most models predict that the
dissolution of marginally bound dSph will not be sudden. Lynden-Bell (1982b)
noted that the major axes of the stellar bodies of many dSph galaxies may
align with their orbits, as predicted for galaxies that are being tidally
distorted.  This effect has been nicely illustrated by McGlynn (1990) in his
Figure 2c.  Since particles which become unbound have low positive energies,
they thus only slowly drift away from their distressed parent galaxy. As
discussed in a slightly different context by Tremaine (1993), a tidally
disrupted object will spread out into a tidal stream of length $s$
determined by the spread in orbital frequency over the size of the object.
After time $t$, the size of the remnant relative to the radius of the
satellite at which tidal stripping began is $s/R_{satellite} \sim \Omega \,
t$, where $\Omega $ is the angular frequency of a circular orbit at the
galactocentric radius where satellite was disrupted. For disruption at 50
kpc, after 1 Gyr the satellite is spread out to only four times its original
size.  As mentioned above, CDM models lead to the expectation of significant
phase-space structure in the Milky Way, remnants of disrupted subsystems,
which may have been observed as `halo moving groups' \eg\ Arnold and Gilmore
(1992). The disruption model for the Galactic dSph system, promoted by Kuhn
(1993), cannot be dismissed out-of-hand, but also does not address the
questions of how these galaxies initially formed, the origins of high {\it
core\/} M/L values, and why they now have such low central stellar
densities. However, it is clear that at least the Sagittarius dwarf and Ursa
Minor show evidence for on-going tidal stretching/disruption, as discussed
above.

An objection to this model that is often raised concerns the supply of dSph,
since unless there is both creation and destruction, we are observing the
dSph at a privileged epoch. However, there may indeed be a source of dSph in
the interaction between the Milky Way and companions.  As noted earlier, the
dSph and the Magellanic Clouds are not uniformly located on the sky, but
occupy two great circles (Kunkel and Demers 1976; Lynden-Bell 1982a,b). This
suggests a tidal-debris origin  -- the break-up of `the Greater Magellanic
Galaxy' and some other now-defunct companion galaxy  -- for the dSph. This
model has several attractive features: (i) It directly addresses the
non-random spatial distribution of the known Galactic dSph. (ii) Objects
that are plausibly self-gravitating dwarf galaxies have been found in the
process of formation in both observed and numerical model tidal-tails
(Barnes and Hernquist 1992; Mirabel, Dottori and Lutz 1992). (iii) This
model naturally gives velocity dispersions for dSph that are close to the 10
km/s HI velocity dispersion of typical galactic disks. (iv) This will yield
low-density galaxies (Gerola, Carnevali and Salpeter 1983).

One has still to explain possible problem areas such as why there are no
dSph of the metallicity of the LMC, the internal kinematics which seems to
require dark matter in the more distant dSph companions of the Milky Way,
and the large range of star formation ages.

\bigskip
\centerline{\bf 5. DWARF SPHEROIDAL GALAXIES BEYOND THE MILKY
WAY}
\medskip
\leftline{\bf 5.1 The Andromeda dSph Galaxies}
\smallskip
The Galactic family of dSph galaxies is not unique.  Van den Bergh
(1972a, 1972b) discovered four candidate dSph galaxies near
the M31 galaxy, one of which (And IV) turned out to be a star cluster in
the M31 disk (Jones 1993).  The remaining three galaxies,  And I,  And II,
and And III, have been shown by Caldwell et al. (1992) to be structurally
similar to Galactic dSph in terms of optical scale lengths, luminosities, and
surface brightnesses (see Figure 4).

The parallel between properties of Galactic and M31 dSph also extends to
stellar population characteristics. These galaxies are like those associated
with the Milky Way in showing no very recent star formation but a range in
stellar ages.  And I is dominated by an old stellar population (Mould and
Kristian 1990), while And II (Aaronson et al. 1985) and And III (Armandroff
et al. 1993) contain AGB stars that are indicators of small intermediate age
(several Gyr) stellar populations.  The improved photometry for And I and
And III of Caldwell et al. provide estimates of the total luminosity that
place the Andomeda dSph on the luminosity-metallicity relationship defined
by the Galactic dSph, as does And II (K\"onig et al. 1993). K\"onig et al.
(1993) also show that there exists significant spread in the metallicity of
the giant branch in And II, as may be expected from the inferred age spread.

By their very existence the Andromeda dSph immediately show that the
dSph cannot be produced by a very rare process.  Even though  surveys for
dSph are still not fully complete around M31,  it evidently is poorer in
dSph than the Milky Way,  despite its larger bulge, close diffuse dwarf E
companions NGC 205 and NGC 185, and ongoing interaction with M32.
That the M31 dSph follow a similar luminosity-metallicity relationship as
the Milky Way systems--albeit for only two objects and with considerable
scatter--also suggests that the evolution of the stellar populations is likely
to be largely controlled by internal processes.  For example, if dSph were
produced by galactic wind sweeping of small companion galaxies or by
the break-up of a larger system, then the metallicity-luminosity
relationship could well differ between host galaxies.  What is not known is
whether the M31 dSph show any evidence for  dense dark matter,
and whether, like their  Milky Way cousins, a range in M/L exists.

\medskip
\leftline{\bf 5.2 Independent dSph Galaxies}
\smallskip

As we move further away from the Milky Way, there is naturally less
information about any class of galaxy. This is an especially severe problem
for the dSph systems due to their small sizes and extremely low surface
brightnesses.  These objects are challenging to find and study as companions
to M31, and indeed even in the extended Milky Way system. However, the
Tucana Dwarf has been suggested to be a distant, isolated dSph by Lavery and
Mighell (1992), who interpreted their color-magnitude diagram as being that
of a dSph galaxy at a distance of about 1 Mpc. These results are confirmed
by deep images taken by Seitzer et al. (1994) with the Wide Field Planetary
Camera 2 on the {\it Hubble Space Telescope}, and thus the Tucana Dwarf
demonstrates that dSph galaxies can exist as independent entities, not
physically associated with any large galaxy.

In addition, there exist fascinating low-luminosity galaxies which have
mixed dIrr/dSph traits. Within the Local Group of galaxies, the dwarf system
LGS3 presents an interesting example of a possible transition object.  LGS3
is not near any giant systems, shows no evidence for HII regions which would
indicate the presence of ongoing large scale star formation (Hunter et al.
1993), but yet contains little HI, $M(HI) \leq 10^6 M_\odot$ (Lo, Sargent
and Young 1993).  Thus LGS3 seems to be a near-dSph, possibly approaching
the stellar fossil phase of its life. Perhaps closer to the dSph systems is
the Phoenix dwarf, which is a low-surface-brightness galaxy with even less
HI (Carignan, Demers, and C\^ot\'e 1991) and a dSph-like stellar population
with a few younger stars included (van de Rydt, Demers, and Kunkel 1991).
The distance estimate of van de Rydt et al. for Phoenix places this galaxy
at a distance comparable to M31, such that it is not bound to a major
galaxy.  However, this estimate is based on correctly identifying the
evolutionary state, and hence luminosity, of resolved red stars, and may
very well be an underestimate. So perhaps we are seeing some evidence that
the final step in becoming a ``pure'' dSph galaxies takes longer in isolated
systems than in dwarfs which are companions to giants, a point which can be
explored as the statistics on faint galaxies in the Local Group improve in
coming years (van den Bergh 1994a,b).

The extensive studies of diffuse dwarf E galaxies in the Virgo cluster also
have begun to give partial answers to these issues (e.g., Binggeli and
Cameron 1991). Photometry of these galaxies suggest that the diffuse dwarf E
 and dSph galaxies may lie on the same surface brightness$-$luminosity
relationship (see Bender et al. 1993), although for such low brightness
galaxies selection effects are a major concern (Phillipps, Davies, and
Disney 1988). In addition both classes of galaxies can be fit with
exponential radial brightness profiles, often have low rotation velocities
for their observed degree of flattening (Bender and Nieto 1990), and likely
have similar luminosity-metallicity correlations. Kormendy and Bender (1994)
use new observations of Virgo Cluster ellipticals to strengthen this
connection. They suggest that dwarf diffuse Es and dSphs are not only a
single structural class, but that this class should be called the dwarf
spheroidals. Sandage and Hoffman (1991) present a related argument for
parallelism between dE and dSph systems in discussing the existence of
Phoenix dwarf-like transition morphology galaxies which sit between the
dwarf irregular and dwarf diffuse E/S0 classes. Physically the dE/dSph
galaxy family can be attributed to similar evolutionary processes in which
mass loss has been a key ingredient (Binggeli 1994).

\medskip

\centerline{\bf 6. CONCLUDING THOUGHTS}
\medskip

Dwarf spheroidal galaxies present a nearby, albeit dim, challenge to current
ideas about the formation and evolution of galaxies. The significance of
these objects will become clearer within the next few years as new
observations yield better information on their most fundamental properties.
An outline of some possibilities is given in Table 3.  We can therefore look
forward to expanding our knowledge of the dSph in several key areas:

1. The ``dark matter issue'' is critical to understanding the true nature of
the dSph galaxies. This issue contains several unresolved questions. First,
we have not yet rigorously shown that the nearer dSph galaxies are in
dynamical equilibrium.  Until we can establish the dynamical states of the
dSph, we will not know if or how much dark matter is required.  If some dSph
systems are indeed now dissolving, then we should also search for ultra- low
surface brightness clouds or moving groups of stars which could mark the
presence of previously disrupted Galactic dSph galaxies (Freeman 1990;
Arnold and Gilmore 1992; Zinn 1993; Majewski 1993).  Indeed, the best
indications that the Galaxy has been caught in the act of eating a companion
galaxy have come from radial velocity surveys (Ibata, Gilmore and Irwin
1994).

At the other extreme, conventional equilibrium dynamical models of the Draco
and Ursa Minor galaxies require very high densities of dark matter.  If the
dark matter in these objects is non-baryonic, then the high densities imply
very early formation times, (see Peebles 1989 for this argument) and the
dSph could well be the oldest class of galaxy. Yet high densities could also
be due to dissipative processes and therefore a sign that this dark matter
is baryonic.  It is too early to draw broad conclusions about the nature of
non-baryonic dark matter on the basis of properties of Galactic dSph
galaxies alone.

2. The non-random distribution of the Galactic retinue of dSph satellites
may be a prime clue of the origins of these systems.  As emphasized by
Lynden-Bell (1982a,b) and more recently by Majewski (1994),
this pattern is suggestive of the production of dSph galaxies
via the tidal disruption of larger systems.  In this model it is hard to see
how the remnants would retain very high densities of non-baryonic dark
matter. We therefore place a high priority on determining if the two great
circles of Galactic dSph systems represents a physical association, e.g.,
through searches for dSph galaxies located elsewhere around the Milky Way
and via proper motion measurements to determine orbital directions for
especially the nearer dSph systems.  As always with the dSphs, nothing is
simple, and the radial velocity data already defy any straightforward model
for the orbital distributions.

3. Some the basic properties of the dSphs, such as metallicity and presence
of an intermediate age population  may correlate with Galactocentric
distance $D_G$ (Silk, Wyse and Shields 1987).
This is puzzling in view of the radial velocity data which
suggest that the dSph are not on high angular momentum orbits that cover
small ranges in $D_G$. The crossing times for these orbits are short and so
we might expect that even if radial gradients were introduced during the
formation of dSph galaxies, they would have been rapidly reduced by orbital
mixing. Better data are needed to define the significance of radial trends
in the properties of Galactic dSph systems, and this should include
comparisons between the properties of Galactic and external dSph
galaxies (Armandroff et al. 1993).

4. Many Galactic dSph satellites have had complicated evolutionary
histories. Although these dSph galaxies have low densities and tiny escape
velocities (even if dark matter is present), they have a range in either
stellar metallicities or ages.  Both features require the host galaxy to
have been influenced by multiple generations of stars, in contrast to the
single stellar generations found in most Galactic globular star clusters.
The kinds of processes responsible for these evolutionary complexities can
be revealed as new measurements define the properties of the main sequence
luminosity functions in the nearest dSph companions to the Milky Way.
Similarly, abundance determinations of iron peak elements, which have an
important source in long-lived Type Ia supernovae, versus elements such as
O, Mg or Si which are predominantly formed in Type II supernovae will
enhance our understanding of the way in which the dSph became polluted to
modest levels with met als.

5. We are in one way indeed observing the dSph at a special epoch. Although
several dSph supported star formation as recently as perhaps 3 Gyr in the
past, none of the dSph are thought to currently be capable of producing new
stars. It is therefore important to understand why these galaxies have
dropped out of the league of star forming galaxies on a time scale that is
only a fraction of the cosmic time. In this regard, the possible existence
or not of a reservoir of `proto-dSph' is a critical issue.  It may be that
this `special epoch' is related to the collapse of bound groups of galaxies
(\cf\ Silk, Wyse and Shields 1987), in which case the existence or not of
isolated dSph is crucial.

6. Study of the satellite galaxies of the Milky Way  will elucidate the
formation and evolution of the Milky Way itself, and hence that of typical
spiral galaxies.  Hierarchical-clustering scenarios of galaxy formation lead
naturally to a synthesis of the ideas presented by Eggen, Lynden-Bell and
Sandage (1962), whereby the stellar halo formed during a rapid, monolithic
collapse phase,  and by Searle and Zinn (1978), whereby an extended more
chaotic accretion phase formed the stellar halo.

ACKNOWLEDGEMENTS

JSG acknowledges partial support from the Graduate School of the University
of Wisconsin and from the Wide Field Planetary Camera 2 Investigation
Definition Team which is supported by NASA through contract NAS7-1260 to the
Jet Propulsion Laboratory; RFGW acknowledges partial support from the NSF
(AST-9016266) and from the Seaver Foundation. The Center for Particle
Astrophysics is supported by the NSF. We thank Kyle Cudworth, Gerry Gilmore,
Ken Freeman, Mario Mateo and Andrea Schweitzer for helpful comments and
discussions.

REFERENCES

\parindent=0pt

Aaronson, M., Gordon, G., Mould, J., Olszewski, E., and Suntzeff, N.
1985, ApJ, 296, L7.

Aaronson, M.  1986, in {\it Stellar Populations}, eds
C.~Norman, A.~Renzini and M.~Tosi (CUP, Cambridge)  p45.

Aguilar, L. A., 1993, in {\it Galaxy Evolution: The Milky Way
Perspective} ed S.~Majewski (ASP, San Francisco) p155.

Aguilar, L. A. and White, S. D. M., 1986, ApJ 307, 97.

Alcock, C., et al., 1993, Nature, 365, 621.

Armandroff, T., Da Costa, G. S., Caldwell, N., and Seitzer, P., 1993,
AJ, 106, 986.

Arnold, R. and Gilmore, G. 1992, MNRAS, 257, 225.

Arp, H. C. and Madore, B.M. 1979, ApJL, 227, L103.

Ashman, K., 1992, PASP, 104, 1109.

Ashman, K., Salucci, P. and Persic, M, 1993, MNRAS, 260, 610.

Aubourg, E.,  et al., 1993, Nature, 365, 623.

Bahcall, J., Flynn, C., Gould, A. and Kirhakos, S., 1994, ApJL in press

Bardeen, J.M., Bond, J. R., Kaiser, N., and Szalay, A. S., ApJ,
1986, 304, 15.

Barnes, J. E. and Hernquist, L. 1992, Nature, 360, 715.

Bender, R. and Nieto, J.-L. 1990, A\&A, 239, 97.

Bender, R., Burstein, D., and Faber, S. M. 1993, ApJ, 411, 153.

Binggeli, B. and Cameron, L. M. 1991, A\&A, 252, 27.

Binggeli, B. 1994, in {\it Panchromatic View of Galaxies: Their
Evolutionary Puzzle}, Ed. G. Hensler, C. Theis, and J. Gallagher
(Editons Fronti\`eres, Gif-sur-Yvette) p172.

Binggeli, B. and Cameron, M. 1991, A\&A, 252, 27.

Binney, J. and Tremaine, S., 1987, {\it Galactic Dynamics} (Princeton
Astrophys., Princeton).

Blumenthal, G.R., Faber, S.M., Primack, J.R. and Rees, M.J., 1984,
 Nature, 311, 517.

Bosma, A., 1982, AJ, 86, 1825.

Byrd, G., Valtonen, M., McCall, M. and Innanen, K., 1994, AJ, 107, 2055.

Capaciolli, M., Piotto, G. and Stiavelli, M. 1993,
MNRAS, 261, 819.

Caldwell, N., Armandroff, T. E., Seitzer, P., and Da Costa, G. S., 1992,
AJ, { 103},  840.

Carignan, C., Demers, S., and C\^ot\'e, S., 1991, ApJL, 381, L13.

Carswell, R.F., Rauch, M., Weymann, R.J., Cooke, A.J. and Webb, J.K., 1994,
MNRAS, 268, L1.

Cen, N. and Ostriker, J. P., 1993, ApJ,  416, 399.

Cudworth, K. M., Olzewski, E. W., and Schommer, R. A. 1986, AJ, 92, 766.

Cudworth, K. M., Schweitzer, A. E., and Majewski, S. R., 1993,  private
communication.

Da Costa, G. S., 1992, in {\it The Stellar Populations of Galaxies}, eds. B.
Barbuy and A. Renzini (Kluwer, Dordrecht), p191.

de Carvalho, R. and Djorgovski, S. 1992,
in {\it Cosmology and Large-Scale Structure in the Universe},
A.S.P. Conference Series, ed R.~de Carvalho (A.S.P., San
Francisco),p135.

Dahn, C. 1994, BAAS, 26, 944.

Dekel, A. and Silk, J., 1986, ApJ, 303, 39.

Demers, S. and Irwin, M. J. 1993, MNRAS, 261, 657.

Demers, S.,  Irwin, M.J. and Gambu, I. 1994, MNRAS, 266, 7.

De Young, D. and Gallagher, J. 1990, ApJL, 356, L15.

Disney, M. J. and Phillipps, S., 1987, Nature, 329, 203.

Djorgovski, S. 1993, in {\it Structure and Dynamics of Globular Clusters},
ASP Conf. Ser. Vol. 50, eds. S. G. Djorgovski and G. Meylan, p373.

Dubath, P., Meylan, G., and Mayer, M., 1992, ApJ, 400, 510.

Eder, J. A., Schombert, J. M., Dekel, A. and Oemler, A., 1989, ApJ,
340, 29.

Eggen, O., Lynden-Bell, D. and Sandage, A., 1962, ApJ, 176, 76.

Evrard, A.E., 1989, ApJ, 341, 26.

Faber, S. M. and Lin, D. C., 1983, ApJL, 266, L17.

Ferguson, H.C. and Sandage, A. 1990, AJ, 100, 1.

Fich, M. and Tremaine, S., 1991, ARAA, 29, 409.

Freedman, W.L., 1994, in `The Local Group', in press...

Freeman, K. C., 1990, in {\it Dynamics and Interactions of Galaxies}, ed.
R. Wielen  (Springer, Berlin), p36.

Freeman, K. C., 1993, private communication.

Gallagher, J. S., 1990, Ann. NY Acad Sci. 596, 1.

Gardiner, L.T., Sawa, T. and Fujimoto, M. 1994, MNRAS, 266, 567.

Gerola, H., Carnevali, P., and Salpeter, E. E., 1983, ApJL 268, L75.

Gilmore, G. and Wyse, R.F.G., 1991, ApJL, 367, L55.

Gunn, J. and Griffin, R. F. 1979, AJ, 84, 752.

Hargreaves, J., Gilmore, G., Irwin, M., and Carter, D., 1994a,
MNRAS, in press.

Hargreaves, J., Gilmore, G., Irwin, M., and Carter, D., 1994b,
MNRAS, submitted.

Hartwick, F.D.A. 1976, ApJ, 209, 418.

Hodge, P. W. 1964, AJ, 69, 438.

Hodge, P.W., 1971, ARAA, 9, 35.

Hodge, P. W., 1982, AJ, 87, 1668.

Hodge P.W. and Michie, R.W., 1969, AJ 74, 587.

Hoessel, J. G., Saha, A., Krist, J., \& Danielson, E. G. 1994, AJ, 108, 645.

Hu, E.M., Huang, J.S., Gilmore, G. and Cowie, L.L., 1994, Nature in press.

Hunter, D.A., Hawley, W. N., and Gallagher, J. S. 1993, AJ, 106, 1797.

Ibata, R., Gilmore, G. and Irwin, M. 1994, Nature, 370, 194

Illingworth, G., 1976, ApJ, 204, 73.

Impey, C., Bothun, G., and Malin, D., 1988, ApJ, 330, 634.

Irwin, M., Bunclark, P., Bridgeland, M. and McMahon, R.J.
1990 MNRAS, 244, 16P

Jones, J. H. 1993, AJ, 105, 933.

Kaiser, N., 1984, ApJL, 284, L9.

Karachentseva, V. E., Schmidt, R., and Richter, G. M., 1984, Astron.
Nachr., 305, 59.

Kaufmann, M. and White, S.D.M., 1993, MNRAS, 261, 921.

Kent, S. M., 1987, AJ, 93, 816.

Knapp, G., Kerr, F., and Bowers, P., 1978, AJ, 83, 360.

K\"onig, C.H.B., Nemec, J.M., Mould, J.R. and Fahlman, G.G. 1993,
AJ, 106, 1819.

Kormendy, J., 1977, ApJ, 218, 333.

Kormendy, J., 1985, ApJ, 295, 73.

Kormendy, J. 1987, in {\it Nearly Normal Galaxies} ed S.M.~Faber
(Springer-Verlag, New York) p163.

Kormendy, J. and Bender, R. 1994, ESO/OHP Workshop on Dwarf
Galaxies, ed. G. Meylan and P. Prugniel, in press.

Kroupa, P., R\"oser, S. and Bastian, U. 1994, MNRAS, 266, 412.

Kuhn, J. R. 1993,  ApJL, 409, L13.

Kuhn, J. R. and Miller, R. H., 1989, ApJL, 341, L41.

Kunkel, W.E. and Demers, S., 1976, Roy. Greenwich Obs. Bull., 182,
241.

Lacey C. and Cole, S. 1993, MNRAS, 262, 627.

Lake, G., 1990a, MNRAS, 244, 701.

Lake, G., 1990b, ApJL, 356, L43.

Larson, R.B., 1974, MNRAS, 169, 229.

Larson, R.  1987, 13th Texas Symposium, eds M.P.
Ulmer (World Scientific) p 426.

Lavery, R. J. and Mighell, K. J., 1992, AJ, 103, 81.

Lee, M. G., Freedman, W., Mateo, M., Thompson, I., Roth, M. and
Ruiz, M.-T., 1993, AJ, 106, 1420.

Lehnert, M. D., Bell, R. A., Hesser, J. E., and
Oke, J. B., 1992, ApJ, 395, 466.

Lin, D. C. and Faber, S. M., 1983, ApJL, 266, L21.

Little, B. and Tremaine, S., 1987, ApJ 320, 493.

Lo, K.Y., Sargent, W.L.W. and Young, K. 1993, AJ, 106, 507.

Lupton, R. H., Fall, M. S., Freeman, K. C., and Elson, R. A. W., 1989,
 ApJ 347, 201.

Lynden-Bell, D., 1982a, Observatory 102, 7.

Lynden-Bell, D., 1982b, Observatory 102, 202.

Majewski, S. 1993, in {\it Galaxy Evolution: The Milky Way
Perspective} ed S.~Majewski (ASP, San Francisco) p5.

Majewski, S. 1994, ApJL, 431, L17.

Majewski, S. R. and Cudworth, K. M., 1993, PASP,
105, 987.

Maloney, P., 1992, ApJL, 398, L89.

Marlowe, A., 1995, PhD thesis, Johns Hopkins University.

Mateo, M., Olszewski, E., Welch, D. L., Fischer, P., and Kunkel, W.,
1991, AJ, 102, 914.

Mateo, M., Olszewski, E. W., Pryor, C., Welch, D. L., and Fisher, P.,
1993, AJ, 105, 510.

Mateo, M., Udalski, A., Szymanski, M., Kaluzny, J., Kubiak, M. and
Krzeminski, W., 1994, preprint.

Mathewson, D., Wayte, S.R., Ford, V.L. and Ruan, K. 1987, Proc Ast Soc
Aust, 7, 19.

Mathieu, R.D., 1985, in  IAU Symposium 113,
{\it Dynamics of Star Clusters}, ed. J.~Goodman and  P.~Hut,  (Reidel,
Dordrecht) p427.

McGlynn, T. A., 1990, ApJ 348, 515.

Meurer, G.A., Freeman, K.C., Dopita, M.A. and Cacciara,
C., 1992, AJ, 103, 60.

Mirabel, I., Dottori, H., and Lutz, D., 1992, A\&A, 256, L19.

Moore, B., Frenk, C.S. and White, S.D.M., 1993, MNRAS, 261, 827.

Mould, J. and Kristian, J. 1990, ApJ, 354, 438.

Mould, J. R., Stavely-Smith, L., and Wright, A. E., 1990, ApJL,
362, L55.

Navarro, J., Frenk, C.S. and White, S.D.M. 1994 MNRAS,267, L1.

Nieto, J.-L., Bender, R., Davoust, E., and Prugniel, P., 1990, A\&A, 230, L17.

Peebles, P.J.E. 1989, in {\it The Epoch of Galaxy Formation}
eds C.S.~Frenk et al.  (Kluwer, Dordrecht) p1.

Phillipps, S., Davies, J. I., and Disney, M. J. 1988, MNRAS, 233, 485.

Pryor, C. and Kormendy, J., 1990, AJ, 100, 127.

Puche, D. and Carignan, C., 1991, ApJ, 378, 487.

Quinn, P. J. and Goodman, J., 1986 ApJ 309, 472.

Rodgers, A. W. and Roberts, W. H. 1994 AJ, 107, 1737.

Sandage, A. 1965, in {\it The Structure and Evolution of Galaxies}, ed.
H. Bondi (Interscience, New York), p.83.

Sandage, A., Binggeli, B., and Tammann, G. A., 1985, AJ, 90, 1681.

Sandage, A. and Hoffman, G. L. 1991, ApJL, 379, L45.

Scholz, R.-D. and Irwin, M. J., 1993, in {\it Astronomy from Wide Field
Imaging} IAU Symposium No. 161, in press.

Schombert, J., Bothun, G., Schneider, S., and McGaugh, S., 1992, AJ, 103,
1107.

Searle, L. and Zinn, R. 1978, ApJ, 225, 357.

Seitzer, P.,  1985, in IAU Symposium 113,
{\it Dynamics of Star Clusters}, ed. J. ~Goodman and P.~Hut,
(Reidel, Dordrecht) p.343

Seitzer, P., Lavery, R. Da Costa, G., Suntzeff, N. and Walker, A.
1994, in preparation.

Silk, J. and  Wyse, R.F.G, 1993, Physics Rep, 231, 293.

Silk, J., Wyse, R.F.G. and Shields, G., 1987, ApJL, 322, L59.

Smecker-Hane, T.A., Stetson, P. B., Hesser, J. E., and
Lehnert, M. D. 1994, AJ, in press.

Songaila, A., Cowie, L. L., Hogan, C. J., and Rugers, M. 1994, Nature,
368, 599.

Suntzeff, N.B. 1993. In {\it The Globular Cluster -- Galaxy Connection},
eds G.~Smith and J.~Brodie (A.S.P., San Francisco) p135.

Suntzeff, N.B., Mateo. M., Terndrup, D.M., Olszewski, E.W., Geisler, D. and
Weller, W., 1993, ApJ, 418, 208.

Thuan, T. X., Gott, J. R., and Schneider, S. E., 1987, ApJL, 315, L93.

Tremaine, S., 1987, in {\it Nearly Normal Galaxies} ed. S. M. Faber
(Springer, New York) p76.

Tremaine, S. 1993, in {\it Back to the Galaxy} eds
F. Verter and S. Holt, (AIP , New York) p599.

Udalski, A. B., et al., 1993, Acta Astronomica, 43, 289.

Vader, J.P., 1986, ApJ, 317, 128.

Vader, P.J. and Sandage, A., 1991, ApJL, 379, L1.

van de Rydt, F., Demers, S., and Kunkel, W. E. 1991, AJ, 102, 130.

van den Bergh, S. 1972a, ApJ, 171, L31.

van den Bergh, S. 1972b, ApJ, 178, L99.

van den Bergh, S. 1992, MNRAS, 255, 25P.

van den Bergh, S. 1994a, AJ, 107, 1328.

van den Bergh, S. 1994b, ApJ, 428, 617.

van der Kruit, P.C. and Searle, L., 1982, A\&A, 110, 61.

van der Kruit P.C. and Shostak, G.S., 1984, A\&A 134,
258.

Walker, T. P., Steigman, G., Schramm, D. N., Olive, K. A., and Kiang,
H.-S., 1991, ApJ, 376, 51.

Wayte, S.R. 1991, in proc IAU Symp 148, The Magellanic Clouds, eds
R.~Haynes and D.~Milne (Reidel, Dordrecht) p15.

Webbink, R. F., 1985, in {\it Dynamics of Star Clusters}, eds J. Goodman and
P. Hut (Reidel, Dordrecht), p541.

Wheeler, J. C., Sneden, C. and Truran, J., 1989, ARAA,
27, 279.

White, S.D.M. 1983, in {\it Morphology and Dynamics of Galaxies},
SAAS-FEE lectures, eds L.~Martinet and M.~Mayor, Geneva
Observatory, p291.

White S.D.M. and Rees, M.J. 1978, MNRAS, 183, 341.

Wirth, A. and Gallagher, J. S., 1984, ApJ, 282, 85.

Wyse, R.F.G. and Gilmore, G., 1992, AJ, 104, 144.

Zaritsky, D., Olszewski, E. W., Schommer, R. A., Peterson, R. S., and
Aaronson, M. A., 1989, ApJ 345, 759.

Zinn, R. 1993, in {\it The Globular Cluster--Galaxy Connection} eds.  G. H.
Smith and
J. P. Brodie, p38.

\vfill\eject
\centerline {FIGURE CAPTIONS}

FIG. 1.--This print is from a IIIaJ plate taken with the Kitt Peak 4-m
Mayall telescope as part of an ongoing astrometric program.
It shows the diffuse
cloud of stars which mark the main body of the Ursa Minor dSph.
The orientation is north up and east to the left; a more detailed
finding image can be found in Cudworth et al. (1986). The dSph
galaxies are difficult to detect when individual stars are unresolved due
to their very low optical surface brightnesses, and yet the optical may
be the best waveband for finding this type of purely stellar galaxy.
(Courtesy of K. Cudworth, Yerkes Observatory)

FIG. 2.--Shown above is  the mean radial stellar density profile for
the Ursa Minor dSph as presented by Hargreaves et al. (1994),
copyright
1994, Royal Astronomical Society; reproduced with permission.
on the
basis of measurements by M. Irwin and D. Hatzidimitriou (in preparation)
with the Cambridge APM.  The solid line is a King model fit which shows
a core and outer tidal cut-off with a possible excess of stars. The
dotted line is an exponential profile fit.

FIG. 3.--The color magnitude diagram in the top panel for the Carina dSph
from Smecker-Hane et al. (1994), copyright 1994, American Astronomical
Society; reproduced with permission. This plot clearly shows a well-defined
giant branch (RGB), an asymptotic giant branch (AGB), horizontal branch (HB)
including the RR Lyrae variables, red clump, and main sequence turnoff. This
type of CMD results from a mixture of old and intermediate age stellar
populations as outlined by the descriptive guide in the lower panel of the
figure.

FIG. 4.--This CCD image
shows the And I dSph.  Structurally And I resembles both the Galactic dSph
systems and fainter, non-nucleated dE galaxies found in the Virgo
Cluster of galaxies. Taken from Caldwell et al. (1992), copyright
1992, American Astronomical Society; reproduced with permission.

\vfill\eject
\tolerance 1500
\singlespace
\parskip=3pt plus 1pt minus 1pt  
  \tabskip=.2em plus .5em minus .2em \baselineskip=12pt
  \def \space{ \noalign{ \vskip4pt \hrule height 1pt \vskip4pt} }

  $$ \vbox{ \halign to \hsize { \hfil#\hfil & \hfil#\hfil &  \hfil#\hfil &
  \hfil# \cr
 \multispan{4} \hfil {Table 1 : {\bf The Local Group }}\hfil\cr
 \noalign{ \vskip 12pt }
 \noalign{ \hrule height 1pt \vskip 1pt \hrule height 1pt }
 \noalign{ \vskip 8pt }
 &  ID &    Type   &  $M_V$ \cr
\space
 Giants:  \cr
 & M31=NGC 224 &   Sb I-II   & $-$21.1 \cr
 & Milky Way   &   SABb/c    & $-$20.6 \cr
 & M33=NGC 598 &   Sc II-III & $-$18.9 \cr
 & LMC         &   Im III-IV & $-$18.1 \cr
\space
 Gas-rich dwarfs:   \cr
 & NGC 6822 &  Im IV-V  & $-$16.4 \cr
 & SMC      &  Im IV-V  & $-$16.2 \cr
 & IC1613   &  Im V     & $-$14.9 \cr
 & WLM      &  Im IV-V  & $-$14.1 \cr
& DDO 210  &  Im V     & $-$11.5 \cr
&  LGS 3    &  dIm/dSph & $-$10.2 \cr
& Phoenix  &  dIm/dSph & $-$9.5  \cr
\space
 Gas--poor dwarfs:   \cr
 & M32=NGC221 & E2       &  $-$16.4 \cr
 & NGC 205    & dE       &  $-$16.3 \cr
 & NGC 185    & dE       &  $-$15.3 \cr
 & NGC 147    & dE       &  $-$15.1 \cr
 &  Fornax     & dSph     &  $-$13.7 \cr
 & Sagittarius & dSph &  $-$13:: \cr
 & And I      & dSph     &  $-$11.8 \cr
 &  And II     & dSph     &  $-$11.8 \cr
& Leo I      & dSph     &  $-$11.7 \cr
& Sculptor   & dSph     &  $-$10.7 \cr
& And III    & dSph     &  $-$10.3 \cr
& Sextans    & dSph     &  $-$10.0 \cr
& Leo  II     & dSph     &  $-$9.9 \cr
& Tucana     & dSph     &   $-$9.5 \cr
& Carina     & dSph     &   $-$9.2 \cr
& Ursa Minor & dSph     &   $-$8.9 \cr
& Draco      & dSph     &   $-$8.5 \cr
 \noalign{ \vskip 8pt\hrule height 1pt \vskip 1pt \hrule height 1pt
  \vskip 8pt }
}}$$
\vfill\eject
  $$ \vbox{ \halign to \hsize { #\hfil & \hfil#\hfil & \hfil#\hfil &
  #\hfil & \hfil#\hfil & \hfil#\hfil & \hfil#\hfil & \hfil#\hfil \cr
 \multispan{8} \hfil {Table 2 : {\bf dSph Properties}} \hfil\cr
 \noalign{ \vskip 12pt }
 \noalign{ \hrule height 1pt \vskip 1pt \hrule height 1pt }
 \noalign{ \vskip 8pt }
Galaxy  &   D$_G$    & M$_V$& $\mu_0$  & $V_r$  & $\sigma_v$    & [Me/H]     &
Age\cr
\space
Carina  &   93      & $-$9.2    & \dots      & 14   & 6    &   $-$1.5    &
Int--Old \cr

Draco   &   75  &  $-$8.5  & \dots    & $-$95  & 9  & $-$1.6 - $-$2.4   & Old
\cr

Fornax &   140  & $-$13.7  & 23.2   & $-$34  & 10   &  $-$1.0 - $-$1.8  &
Int--Old \cr

Leo I  &   270 & $-$11.7  & 22.3  & 177   & \dots    & $-$1.3    & Int--Old
  \cr

Leo II    & 215 & $-$9.9  & 23.8:  & 16   & \dots        & $-$1.9 & Old: \cr

Sagittarius &  12::   & $ -$13:: & 25.4  &  176 & \dots  & $-$1.2 & Int    \cr

Sculptor  & 79 & $-$10.7  & 24.1   & 74   & 7   & $-$1.5 - $-$2.0  & Old \cr

Sextans   & 85 & $-$10.0  & 25.5   & 78   & 6   & $-$1.7 - $-$2.5 & \dots \cr

Ursa Minor & 65  & $-$8.9  & 25.1  & $-$88  & 11     &  $-$2.2      & Old \cr
 \noalign{ \vskip 8pt\hrule height 1pt \vskip 1pt \hrule height 1pt
  \vskip 8pt }
}}$$

\doublespace

Notes: D$_G$ is the Galactocentric distance in kpc,        $\mu_0$ is
central surface brightness in  V mag/square arcsec, $V_r$ is the
Galactocentric radial velocity in km/s and $\sigma_v$ is the observed
stellar velocity dispersion in km/s. Distances and radial velocities are
from  Zaritsky et al. (1989), except for Leo II (Lee et al. 1993),
Sagittarius (Ibata, Gilmore and Irwin  1994), and Sextans (Irwin et al.
1990; da Costa et al. 1991).  The M$_V$ are from van den Bergh (1992), with
the exceptions of the Sagittarius dwarf (Ibata, Gilmore and Irwin 1994) and
Leo II (Demers and Irwin 1993); small differences are found in tabluations
of M$_V$ but given the uncertainties of typically 0.3 mag for measurements
of these low-surface-brightness systems (e.g., see Hodge 1982; da Costa et
al. 1991), these dominate over most of the stated distance uncertainties.
The surface brightness of the Sagittarius dwarf is taken from Mateo \etal\
(1994). Abundances are from the literature and include results from
spectroscopy of individual stars (e.g., Lehnert et al. 1992) and derivations
based on giant branch properties (e.g., Lee et al. 1993). Stellar population
ages are our assessments based on the current literature where stellar
populations with ages of $\leq$ 10 Gyr are classified as intermediate age.

\singlespace

\vfill\eject

  $$ \vbox{ \halign to \hsize { #\hfil & \hfil#\hfil & \hfil#\hfil &
\hfil#\hfil & \hfil#\hfil & \hfil#\hfil & \hfil#\hfil \cr
\multispan{7}
\hfil {Table 3 : {\bf Important Observational Programs }} \hfil\cr
\noalign{ \vskip 12pt }  \noalign{ \hrule height 1pt \vskip 1pt \hrule
height 1pt }  \noalign{ \vskip 8pt }
Objective       & Surveys  &  Deep Ptm
& Multi-Wavelength & Spectra    &    & Proper Motion \cr
& & & & Hi-Res & Lo-Res \cr
\space 1. More dSphs        & C  & & I \cr
2. Radial Profiles & & C & & & I \cr
3. Kinematics  & & & & C & I & C \cr
4. Gas Content & & C & C \cr
5. Metallicity & & I & I & C & I \cr
6. Stellar Ages & & C & I & I & C & I                   \cr
7. Orbits  & I & I & & I & I & C \cr
 \noalign{ \vskip 8pt\hrule height 1pt \vskip 1pt \hrule height 1pt
  \vskip 8pt }
}}$$

\doublespace
Explanation and notes:

Table entries refer to important (I) or critical (C) observations. Deep Ptm
includes imaging to faint levels such as is planned with the HST, Surveys
are wide angle studies, Multi- Wavlength includes radio, radio, infrared,
VUV, and other types of measurements, Spectra are high and low spectral
resolution for stars, and Prop Motion are absolute proper motion
determinations.

The objectives are (1) Find more examples of dSph galaxies around the Milky
Way and in the nearby universe; (2) Measure radial stellar surface density
profiles and shapes; (3) Determine internal radial velocity distribution
functions versus radius; (4) Undertake surveys for ionized, neutral and
molecular gas, including circumstellar ejecta; (5) Stellar metallicity
determinations, including abundance ratios for key elements which are
sensitive to the nucleosynthesis history; (6) Ages of stars; and (7) Orbits
for individual Galactic dSph systems.

\bye